\pdfoutput=1

\documentclass[conference]{IEEEtran}
\usepackage{float}
\usepackage{cite}
\usepackage{amsmath}
\usepackage[T1]{fontenc}
\usepackage[utf8]{inputenc}
\usepackage{mathtools}  
\usepackage{amssymb}

\makeatletter
\renewcommand{\fnum@figure}{Figure \thefigure}
\makeatother

\ifCLASSINFOpdf
  
\else
  
\fi

\begin{document}

\title{Bare Demo of IEEEtran.cls\\ for IEEE Conferences}

\title{Quantitative Control Approach for Wind Turbine Generators to Provide Fast Frequency Response with Guarantee of Rotor Security}

\author{\IEEEauthorblockN{Siqi Wang, Kevin Tomsovic}
\IEEEauthorblockA{Dept. of Electrical Engineering and Computer Science\\
University of Tennessee\\
Knoxville, TN 37996\\
Email: swang31@utk.edu}\\[-4.0ex]
}

\maketitle

\begin{abstract}
Wind generation is expected to reach substantially higher levels of penetration in the near future. With the converter interface, the rotor inertia of doubly-fed induction generator (DFIG) based wind turbine generator is effectively decoupled from the system, causing a reduction in inertial response. This can be compensated by enabling the DFIG to provide fast frequency response. This paper proposes a quantitative control approach for DFIG to deliver fast frequency response in the inertial response time scale. A supplementary power surge function is added to the active power reference of DFIG. The exact amount of power surge that is available from DFIG-based wind turbine is quantified based on estimation of maximum extractable energy. Moreover, the operational constraints such as rotor limits and converter over-load limit are considered at the same time. Thus, the proposed approach not only provides adequate inertial response but also ensures the rotor speed is kept within a specified operating range. Rotor safety is guaranteed without the need for an additional rotor speed protection scheme. 
\end{abstract}


\IEEEpeerreviewmaketitle

\section{Introduction}
In recent years, the installed capacity of wind power has grown significantly. One major concern of a high penetration of wind generation is the reduced inertia, which results in deterioration of inertial response. According to\cite{6344717} and\cite{5275247}, Electric Reliability Council of Texas (ERCOT) and  Western Interconnect (WECC) have both reported deficiency in inertial response due to an increase of converter interfaced wind resources. Conventionally, synchronous generators serve as the major contributor to provide inertial response. With more synchronous generators forced offline, the burden could be shifted to the wind turbine generators as well. In fact, the speed change of synchronous generators is usually limited to between 1.0 $p.u$ to 0.95 $p.u$ and only 9.75\% of the kinetic energy can be drawn, but asynchronously coupled DFIG can allow its speed to drop from 1.0 $p.u$ to 0.7 $p.u.$, so the potential kinetic energy drawn from DFIG could be much greater than that from conventional generators \cite{Banakar2009}. 

To release the potential kinetic energy from DFIG, a general approach is to introduce a support function as a supplementary control loop, adding to the active power reference. The existing approach for DFIG to provide inertial response can be roughly divided into three categories characterized by the shape of the support function: $df/dt$ response; $\Delta f$ response; and fixed trajectory response \cite{GainScheduling}. The work reported in \cite{Inertia2} introduced an additional supplementary controller proportional to the derivative of system frequency to emulate the natural inertial response ($df/dt$ response). Similarly, the work in \cite{Inertia1} modifies the supplementary control signal to be proportional to the frequency deviation to achieve a greater contribution from the wind turbines ($\Delta f$ response). However, both $df/dt$ and $\Delta f$ response are scenario-based, the response shape depends on the system frequency measurement $f$ and the control gains. Moreover, $df/dt$ is sensitive to measurement error and noise \cite{Inertia1}. The shape of the fixed trajectory is not determined by the frequency disturbance but is based on a power surge function. The power surge function can be various types, such as the kinetic discharge function shown in \cite{Banakar2009} and the temporary frequency control function (TFCF) shown in \cite{TPI}. 

However, the following two questions remain open regarding the fixed trajectory response approach: 

1) How to quantitatively design the power surge function to maximize a DFIG's contribution without stalling the rotor? 

2) How to coordinate multiple DFIGs under different operating conditions to provide fast and adequate system response?

This paper investigates these two questions and proposes a possible solution - a quantitative control approach for DFIG to provide fast frequency response with guaranteed rotor security. First, the maximum power injection for a given DFIG is identified based on wind turbine maximum extractable energy and the minimum allowable rotor speed. Then a coordination scheme is presented for DFIGs with different wind speed to deliver an adequate response. 

The rest of the paper is organized as follows: Section \uppercase\expandafter{\romannumeral2} introduces the DFIG modeling approach and the control system. Section \uppercase\expandafter{\romannumeral3} presents the proposed control approach. Case studies are demonstrated on a 181-WECC system with 50\% wind penetration in Section \uppercase\expandafter{\romannumeral4}. Conclusions are given in Section \uppercase\expandafter{\romannumeral5}.

\section{DFIG Model and Control System}
\begin{figure}[!htbp]
\centering
\includegraphics[width=0.50\textwidth, height=0.23\textwidth]{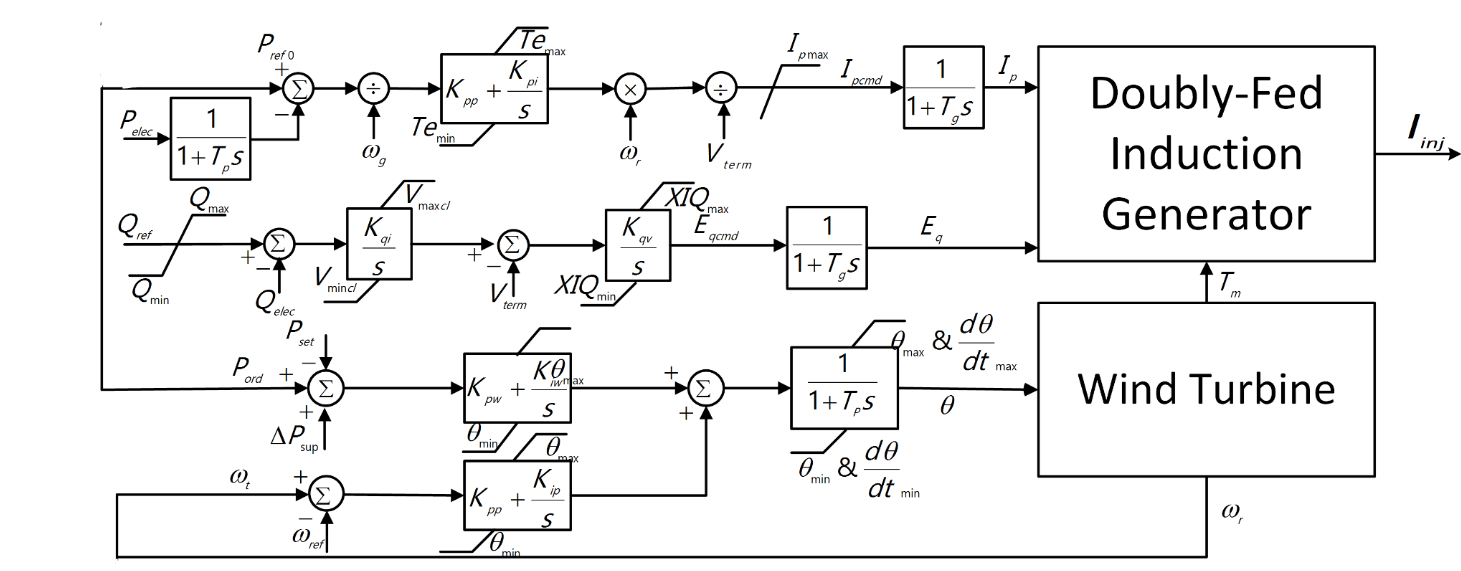}
\caption{DFIG-based wind turbine generator model}
\label{DFIG-wt}
\end{figure}

The objective of this paper is to study the large-scale power system frequency response, so the positive sequence type-3 DFIG machine model is adopted here. The generator is modeled as a controlled current source, while the decoupled active current $I_{p}$ and excitation current $I_{q}$ determine the required active and reactive power injection to the grid \cite{type-3}. The basic DFIG-based wind turbine generator model are shown in Figure \ref{DFIG-wt}. 

Two first-order low-pass filters with a time constant of $T_{g} = 20$ ms will represent the converter control systems \cite{wind_AGC}. This model has been validated and widely used in large-scale system wind integration studies in previous studies, such as, \cite{Inertia1} and \cite{wind_application}. It is assumed that DFIGs within the same wind farm has the same wind speed and can be aggregated into a single unit with the same MVA rating as the summation of each DFIG's MVA rating \cite{Inertia1}. The wind speed spatial variation across regions is not the focus of this work, and thus, is neglected. A similar assumption can be found in \cite{Inertia1} and \cite{deload_pitch}. The regular operation mode for DFIG is based on the widely used maximum power point tracking (MPPT) strategy. The active power reference of DFIG for a measured rotor speed $\omega_{r}$ is:
\begin{equation}\label{MPPT}
P^{mppt}_{ref} =  
\begin{cases*}
k_{opt}\omega_{r}^{3} , & if $\omega_{r} < \omega_{r}^{max}$ \\
P_{max}, & if $\omega_{r} \geq \omega_{r}^{max}$
\end{cases*}
\end{equation}
with
\begin{equation}
k_{opt} = \frac{1}{2} \rho A \frac{R^{3}}{\lambda^{3}}C_{p}(\lambda_{opt,\beta=0},0)
\end{equation}
where $\rho$ is the air density, $A$ is the wind turbine swept area, $R$ is the wind turbine radius, $C_{p}$ is the power coefficient, $\beta$ is the pitch angle, $\lambda$ is the tip speed ratio and $V_{w}$ is the wind speed \cite{Pulgar}.

The mechanical power $P_{m}$ extracted from a wind turbine can be expressed as :
\begin{equation}\label{Pm}
P_{m} = \frac{1}{2}\rho A C_{p}(\lambda,\beta) V_{w}^{3}
\end{equation}
The one-mass drive train model with wind turbine inertia $H_{t}$ is:
\begin{equation}
\frac{d\omega_{r}}{dt} = \frac{1}{2H_{t}}(P_{m}-P_{e})
\end{equation}

\section{Fast Frequency Response from DFIG with Consideration of Rotor Limits}
During normal operation, DFIG still follows the power reference of MPPT as shown in (\ref{MPPT}). During frequency disturbances, the MPPT mode is bypassed, the support mode is activated to inject excess active power $\Delta P$ from wind turbine and the $\Delta P$ is added to the initial power reference $P_{ref}^{mppt}$. Due to fast response of converter control action, the electrical power output will quickly increase to $P_{ref}^{mppt}+\Delta P$, but turbine aerodynamic power will remain near $P_{m}$. The mismatch will force rotor speed to decelerate and the stored kinetic energy will be released. With no reserve allocated, the DFIG on support mode is expected to only provide temporary frequency support, the active power output will finally return to maximum power point, and thus, the overall response should include a short over-production period and a recovery period to allow for rotor speed restoration as shown in Figure \ref{TPI}. In this work, we propose to calculate the exact amount of power injection $\Delta P_{max}$ based on estimated maximum deliverable energy $\Delta E_{del}$, maximum rotor speed deviation $\Delta \omega_{rmax}$ and maximum converter overload limit $P_{lim}$.
\subsection{Estimation of Maximum Power Injection $\Delta P_{max}$}
From the energy point of view, the energy balance of DFIG in support mode can be expressed as follows \cite{MaxEnergyThesis}:
\begin{equation}\label{Etotal}
\Delta E_{aero} - \Delta E_{e} - \Delta E_{loss} - \Delta E_{kic} = 0
\end{equation}
where
$\Delta E_{aero}$ is the aerodynamic energy variation, $\Delta E_{e}$ is the electric energy variation, $\Delta E_{loss}$ is the energy losses, $\Delta E_{kic}$ is the kinetic energy variation, $T_{del}$ is the total delivery time. 
Assuming wind speed is constant during a temporary power injection period:
\begin{equation}\label{dE}
\Delta E_{aero} = \int_{0}^{T_{del}}\Delta P_{m}dt = KV^{3}\int_{0}^{T_{del}}\Delta C_{p}dt
\end{equation}
Approximating $\Delta Cp$ as the first two terms of Taylor’s series yields \cite{MaxEnergyThesis}:
\begin{equation}\label{dCp}
\Delta C_{p} = \frac{\partial C_{p}}{\partial t}|_{t=0}t + \frac{\partial^{2} C_{p}}{\partial t^{2}}|_{t=0}\frac{t^{2}}{2}
\end{equation}
Further assuming the rate of rotor speed deviation is constant, while $d\omega_{r}/dt $ is constant, the partial derivative of the power coefficient in (\ref{dCp}) can be approximated as:
\begin{equation}\label{pCp1}
\begin{aligned}
\frac{\partial C_{p}}{\partial t}|_{t=0} &= \frac{\partial C_{p}}{\partial\lambda}\frac{\partial\lambda}{\partial\omega}\frac{\partial\omega}{\partial t}|_{t=0}=\frac{R}{V}\frac{\partial C_{p}}{\partial\lambda}\frac{\partial\omega}{\partial t}|_{t=0}\\
\frac{\partial^{2} C_{p}}{\partial t^{2}}|_{t=0} &=(\frac{R}{V})^{2} \frac{\partial^{2} C_{p}}{\partial\lambda^{2}}(\frac{\partial\omega}{\partial t})^{2}|_{t=0}+\frac{R}{V}\frac{\partial C_{p}}{\partial\lambda}\frac{\partial^{2}\omega}{\partial t^{2}}|_{t=0}
\end{aligned}
\end{equation}
Integrating (\ref{dE}), (\ref{dCp}) and (\ref{pCp1}) over time $T_{del}$ yields:
\begin{equation}\label{Eaero}
\begin{split}
\Delta E_{aero} & = \frac{1}{2}KRV^2\frac{\partial C_{p}}{\partial\lambda}|_{t=0}T_{del}\Delta\omega_{r}\\
& +\frac{1}{6}KR^{2}V\frac{\partial^{2} C_{p}}{\partial\lambda^{2}}|_{t=0}T_{del}\Delta\omega^{2}_{r}
\end{split}
\end{equation}
The kinetic energy variation during the support period is:
\begin{equation}\label{Ekic}
\Delta E_{kic} = \frac{1}{2}J(\omega^{2}_{r2}-\omega^{2}_{r1})
\end{equation}
and the electric energy variation is:
\begin{equation}\label{Ee}
\Delta E_{e} = \int_{0}^{T_{del}}\Delta P_{del}dt = \Delta E_{del}
\end{equation}
The energy loss variation can be approximated by mechanical losses as:
\begin{equation}\label{Eloss}
\Delta E_{loss} = (1-\eta)\Delta E_{aero} 
\end{equation}
where $\eta$ is the energy efficient.
Thus, combining (\ref{Etotal}), (\ref{Eaero}), (\ref{Ekic}), (\ref{Ee}) and (\ref{Eloss}), the overall energy delivered during power injection period can be expressed as follows:
\begin{equation}\label{E_del_1}
\begin{split}
\Delta E_{del} & = (\frac{\eta}{12} K V_{w}^{2} \frac{\partial C_{p}}{\partial\lambda}|_{t=0} -J \frac{\omega_{r}(0)}{\omega_{s}^{2}} T_{del})\Delta\omega_{r}\\
& +(\frac{\eta}{36}KV_{w}^{2}\frac{\partial C_{p}}{\partial\lambda}|_{t=0}+\frac{\eta}{216}KV_{w}\frac{\partial^{2} C_{p}}{\partial\lambda}^{2}|_{t=0}\\
& -\frac{J}{2\omega^{2}_{s}}T_{del})\Delta\omega^{2}_{r}
\end{split}
\end{equation}
This expression reveals the relationship between the delivered energy $\Delta E_{del}$, the rotor speed deviation $\Delta\omega_{r}$ and the delivery time $T_{del}$. By specifying $T_{del}$ and allowable maximum rotor speed deviation $\Delta\omega_{r}^{max}$, the maximum extracted energy and the corresponding power injection $\Delta P_{del}$ can be calculated. 

\begin{figure}[h]
	\centering
	\begin{minipage}[]{0.22\textwidth}
		\centering
		\includegraphics[width=\textwidth]{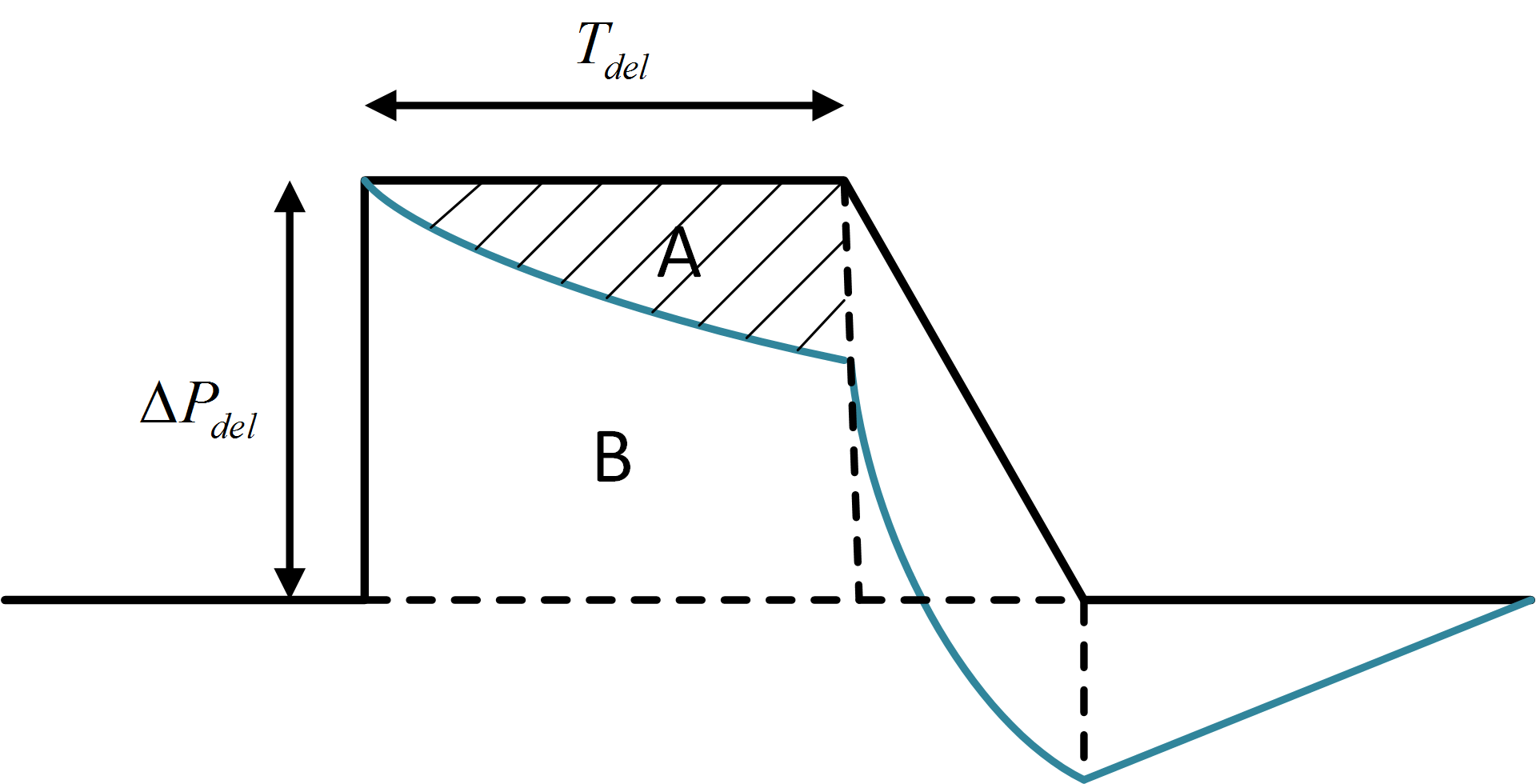}
		\caption{Illustrative example of conservatives}
		\label{TPI}
	\end{minipage}\hfill
	\begin{minipage}[]{0.24\textwidth}
		\centering
		\includegraphics[width=\textwidth]{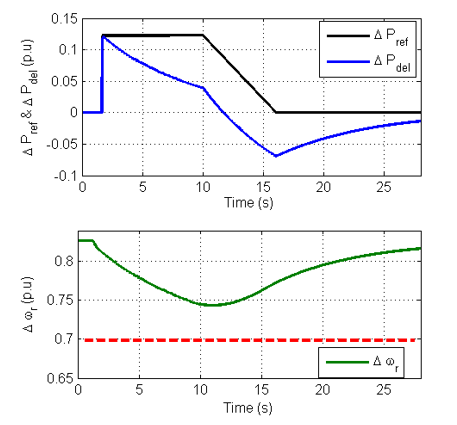}
		\caption{DFIG response example}
		\label{rotorspeed}
	\end{minipage}\hfill
\end{figure}

Note when applying the calculated maximum power injection, the actual energy released by DFIG is smaller than calculated. As shown in Figure \ref{TPI}, the supplementary power reference is represented by the black curve, but the true response curve is shown in blue. The response reduces the rotor speed (the $k_{opt}\omega_{r}^{3}$ term in (\ref{MPPT})), and thus, reduces the power reference and tapers off the ideal response\cite{ideal_Inertial}. A similar observation is obtained in Ref \cite{ideal_Inertial}. While the calculation is conservative, this can guarantee the rotor speed of DFIG always remains higher than the minimum allowable limit, and no additional rotor speed protection scheme is required. 

A DFIG response example with $V_{w}$ $=$ 9.44 $m/s$ and minimum allowable rotor speed $\omega_{r}^{min}$ $=$ 0.7 $p.u$ is shown in Figure \ref{rotorspeed} to illustrate the conservative nature of the result.

Another constraint that must be taken into consideration when calculating the maximum active power injection is the converter limit (here assumed to be $P_{lim} = 1.2$ p.u). Finally, the maximum power injection of a DFIG on support mode at a given wind speed $V_{w}$ can be expressed as:
\begin{equation}
\Delta P_{max}(V_{w}) = \min(\Delta P_{del},\,P_{lim} - P_{0}(V_{w}))
\end{equation}
where $P_{0}(V_{w})$ is the initial operating point of DFIG with wind speed $V_{w}$. The $P_{max}(V_{w})$ can be saved as a look-up table and used as criteria for on-line selection of DFIGs for delivery of fast frequency response.

\subsection{Sensitivity Measure of Frequency Nadir to Active Power Change}
\begin{figure}[h]
	\centering
	\begin{minipage}[]{0.24\textwidth}
		\centering
		\includegraphics[width=\textwidth]{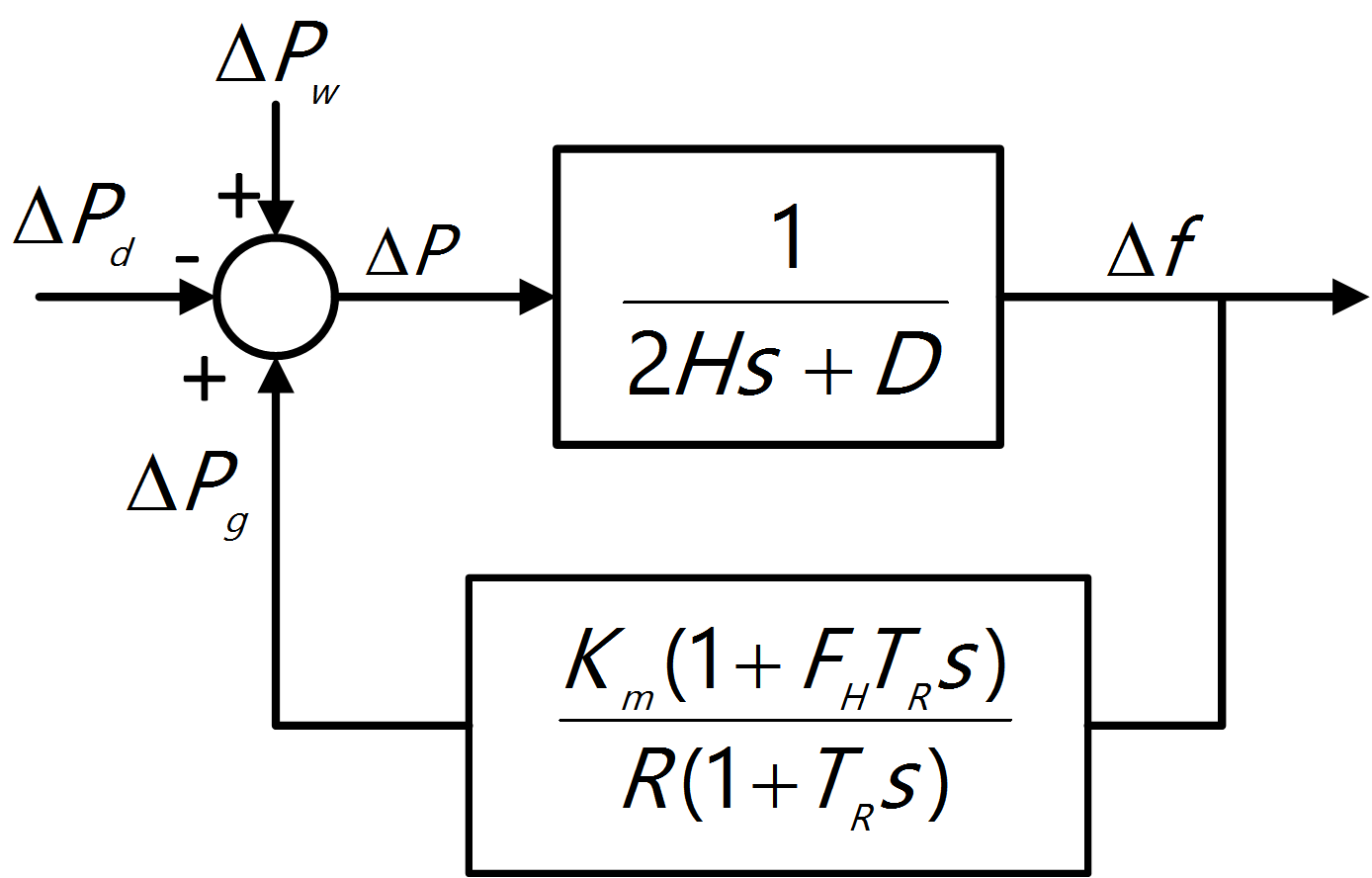}
		\caption{LFR model.}
		\label{LFR model}
	\end{minipage}\hfill
	\begin{minipage}[]{0.22\textwidth}
		\centering
		\includegraphics[width=\textwidth]{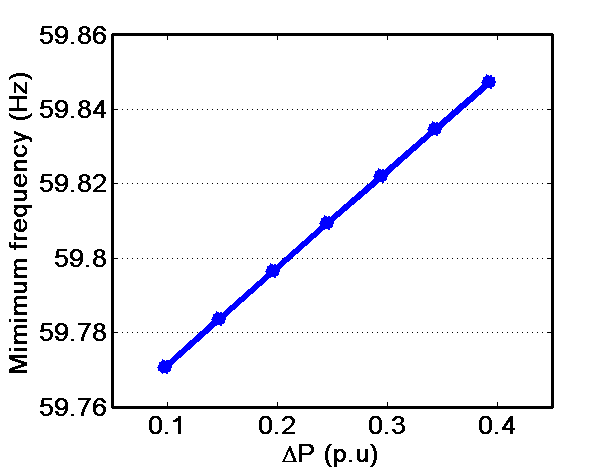}
		\caption{$f_{min}$ varies with $\Delta P$.}
		\label{fnadir}
	\end{minipage}\hfill
\end{figure}

The second step is to obtain the sensitivity measure of frequency nadir variation to active power injection change. Consider, the first order frequency response model shown in Figure \ref{LFR model}, the frequency nadir of this system is given by \cite{fnadir}:
\begin{equation}\label{loworder}
f_{min} = f_{b}(1-\frac{R}{K_{m}+DR}[1+\alpha e^{-\xi\omega_{n} t_{n}sin(\omega_{t} t_{n}+\phi)}]\Delta P)
\end{equation}
with 
\begin{equation}\label{notation}
\begin{aligned}
\omega^{2}_{n} &= \frac{DR+K_{m}}{2HRT_{R}}\\
\xi &= \frac{2HR+(DR+F_{H})T_{R}}{2(DR+K_{m})}\omega_{n}\\
\alpha &=\sqrt{\frac{1-2T_{R}\xi\omega_{n}+T^{2}_{R}\omega^{2}_{n}}{1-\xi^{{2}}}}\\
\omega_{r} &=\omega_{n}\sqrt{1-\xi^{2}}\\
\phi &= tan^{-1}(\frac{\omega_{r}T_{R}}{1-\xi\omega_{n}T_{R}}+ tan^{-1}(\frac{\sqrt{1-\xi^2}}{\xi}))
\end{aligned}
\end{equation}
where $f_{b}$ is the base frequency, $K_{m}$ is the mechanical power gain factor, $R$ is the droop gain, $D$ is the damping coefficient. $T_{R}$ is the reheat time constant and $F_{H}$ is the fraction of total power generated by the HP turbine. The detailed explanation of (\ref{loworder}) and (\ref{notation}) with the notations can be found in \cite{loworder} which is not discussed in detail in this paper.

Equation (\ref{loworder}) shows that the overall injected active power $\Delta P_{max}$ to the system has a near linear relationship with the system frequency nadir $f_{min}$. A similar conclusion is reached in Ref \cite{Commitment_Load}, and is further verified by a detailed transient simulation result of a nonlinear power system in this paper as shown in Figure \ref{fnadir}.

The above analysis indicates the sensitivity measure of frequency nadir to a real power change can be assessed by the matrix $S = \partial\Delta f_{min}/\partial\Delta P_{max}$ for all participating DFIGs with different wind speed using offline simulations. With the $\Delta P_{max}(V_{w})$ and the matrix $S$, the minimum commitment of DFIGs to avoid excessive frequency excursions from $f_{min}$ to $f_{min}+\Delta f_{min}$ can be computed for each possible under-frequency event offline. During online operation, when detecting the under-frequency event, the scheduled minimum number of DFIGs on support mode will be initiated to ensure the frequency nadir will be above the specified threshold.

\subsection{Summary of the approach}
\begin{figure}[!htbp]
\centering
\includegraphics[width=0.4\textwidth,height=0.3\textwidth]{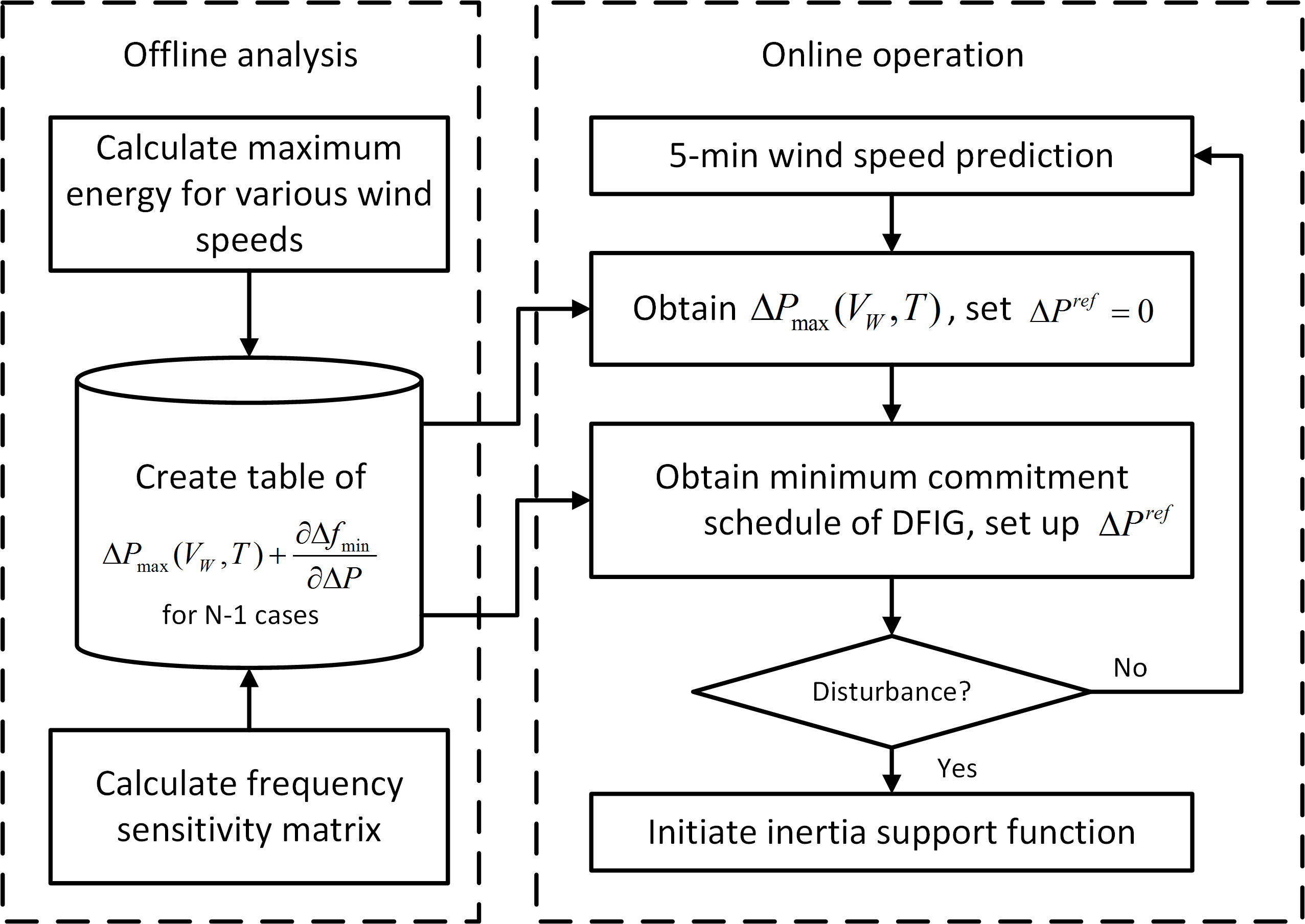}
\caption{Flowchart of proposed control approach.}
\label{flowchat}
\end{figure}

The flow chart of the proposed approach is shown in Figure \ref{flowchat}.
It is assumed that the short term wind speed prediction is available every 5 minutes. The sensitivity matrix $S$ and maximum power injection $\Delta P_{max}(V_{w})$ can be obtained and stored in a database. The minimum number of DFIGs on support mode that need to provide fast frequency response is scheduled at the beginning of each 5-minute interval. Once detecting a frequency deviation out of the specified dead-band, the on-line process activates the scheduled DFIGs. The power reference deviation is then calculated and added to the MPPT reference.

\section{Case Studies and Simulation Results}
A DFIG based wind farm model is developed as a user-defined model in OpenModelica, and compiled as a functional mock-up unit (FMU) then linked to ePHASORsim, which is a real-time power system dynamic simulation software \cite{ephasorsim}. The proposed control approach is implemented on the 181-bus WECC system presented in \cite{181_bus} and modified to have 50\% wind penetration. The wind locations are chosen based on NREL wind integration database which is available in \cite{NREL}. The other synchronous units in the system were modeled with a round rotor machine (GENROU), AC exciter (IEEEX1), turbine governor (IEEEG1) and a power system stabilizer (IEE2ST). To represent the equivalent frequency of the system, a center of inertia (COI) reference frequency is used as the monitoring frequency for comparison of different scenarios \cite{COI}. 

\begin{figure}[!htbp]
\centering
\centering\includegraphics[width=0.48\textwidth,height=0.24\textwidth]{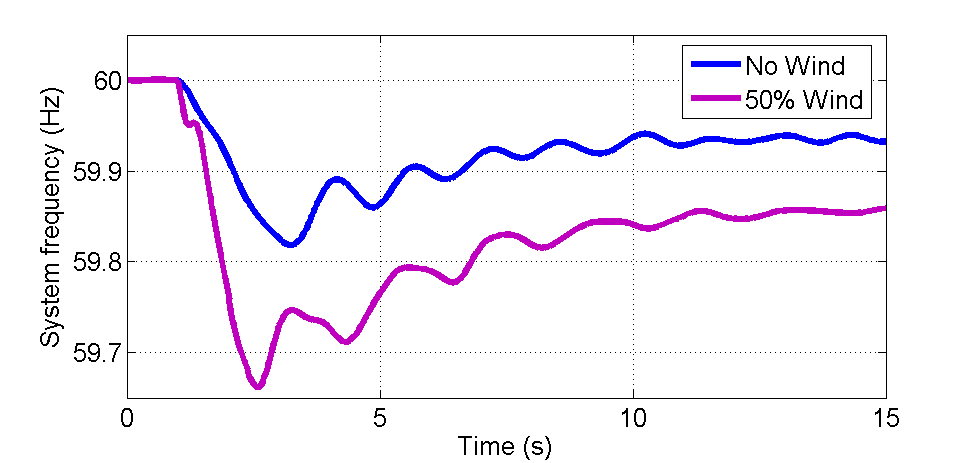}
\caption{Frequency response comparison.}
\label{f_50}
\end{figure}

In the first case study, the largest generator trip event (2160 MW) is evaluated for the studied system. As shown in Figure \ref{f_50}, the blue curve is the response of the studied system without wind penetration under this event, the purple curve is the frequency response to the same event with 50\% wind penetration while all DFIGs are on MPPT with no contribution to frequency disturbance.
Referring to (\ref{loworder}) in the previous section, the replacement of the synchronous generators reduces the overall equivalent inertia constant, leads to a higher ROCOF, and causes the frequency nadir to decrease to 59.66 Hz from 59.81 Hz which is close to the under frequency load shedding limit (59.65 Hz). At the same time, the less effective governor response deteriorates the primary regulation capability, reducing the settling frequency from 59.94 Hz to 59.86 Hz. 

Next, five DFIG units with wind speed ranging from 7 $m/s$ to 12 $m/s$ are scheduled to provide frequency support with the same system condition as previous. The matrix $S$ and $\Delta P_{max}$ for each wind speed are available in the database and are given in Table \ref{table1}.
\begin{table}[h!]
\centering
\caption{$\partial \Delta f_{min}/\partial \Delta P_{max}$, $\Delta P_{max}$ under various $V_{w}$}
\begin{tabular}{c|c|c|c} 
\hline
Unit Number & $V_{w}(m/s)$ & $\Delta P_{max} (p.u)$ & $\partial \Delta f_{min}/\partial \Delta P_{max} (Hz/pu)$ \\
\hline
DFIG unit 1 & 10.35 & 0.3347 & 0.1526 \\ 
DFIG unit 2 &  9.83 & 0.2652 & 0.2089  \\
DFIG unit 3 &  7.90 & 0.0754 & 0.1259 \\
DFIG unit 4 & 11.00 & 0.4339 & 0.0686  \\
DFIG unit 5 &  8.50 & 0.1221 & 0.2530 \\
\hline
\end{tabular}
\label{table1}
\end{table}

\begin{figure}[!htbp]
\centering
\centering\includegraphics[width=0.46\textwidth,height = 0.26\textwidth]{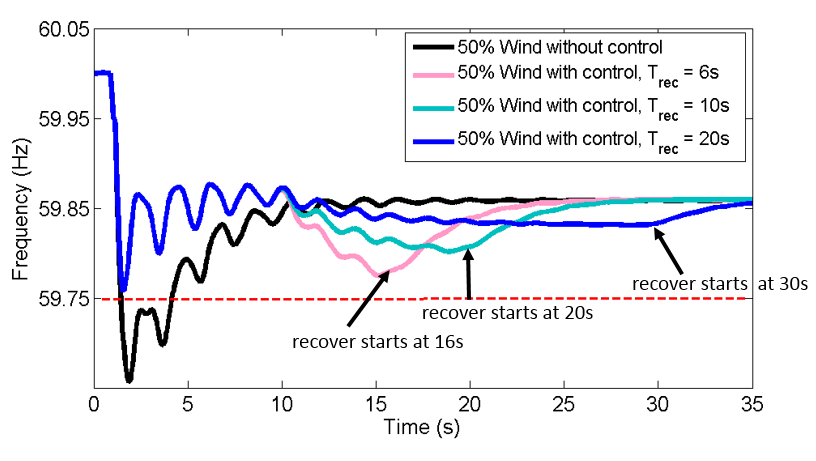}
\caption{Frequency response comparison under MPPT and support mode}
\label{f_TPI}
\end{figure}

To ensure a 0.1 Hz security margin from the UFLS threshold, the frequency nadir threshold is set to be 59.75 Hz. Considering the same worst-case generator trip event as previous case study, when the DFIG nearby bus frequency is beyond the dead-band, the minimum number of DFIG units on support mode are switched in. The frequency response is shown in Figure \ref{f_TPI}. 

In this particular case, two units (DFIG unit 1 and DFIG unit 2 in Table I) are required to meet the 0.1 Hz frequency nadir improvement. Compared to the no control case (the black curve in Figure \ref{f_TPI}), the initial ROCOF is smaller and the initial frequency nadir is above the specified 59.75 Hz. It is also noticeable that there is a second frequency dip associated with the fact that the rotor speed of the DFIG needs to recover back to pre-disturbance speed. This can only be avoided if the DFIG is on reserve mode or compensated by other resources. Still, it can be improved by adjusting recovery time. Figure \ref{f_TPI} also shows the frequency response with different recovery time. It can be seen that when increasing the recovery time, the frequency dip is reduced but the recovery time is prolonged. System planners must ensure that the manner in which energy is delivered and recovered is optimal for the specific system \cite{GainScheduling}. 

\begin{figure}[!htbp]
\centering
\centering\includegraphics[width=0.48\textwidth,height = 0.24\textwidth]{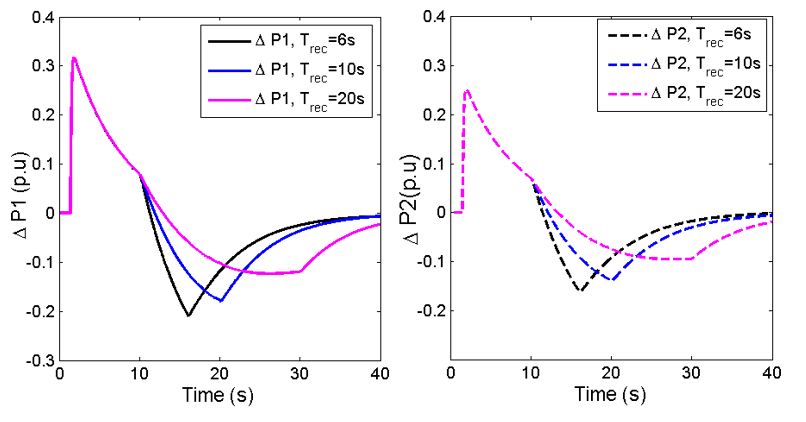}
\caption{Active power of scheduled DFIGs under support mode.}
\label{P_TPI}
\end{figure}

\begin{figure}[!htbp]
\centering
\centering\includegraphics[width=0.48\textwidth,height = 0.24\textwidth]{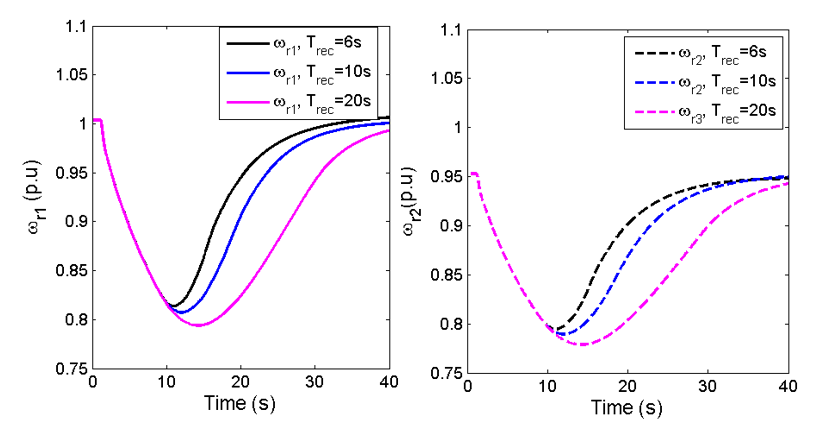}
\caption{Rotor speed of DFIGs under support mode.}
\label{wr_TPI}
\end{figure}

Figure \ref{P_TPI} shows the active power contribution of each DFIG unit, which indicates the maximum power injection of each DFIG unit follows the schedule. The rotor speed response for each DFIG unit is shown in Figure \ref{wr_TPI}, and verifies that the rotor speed remains above the minimum allowable limit.
\section{Conclusion}
This paper presents a quantitative control approach for DFIG to provide fast frequency response by incorporating a supplementary power surge function to the existing MPPT reference. The supplementary power surge amount is carefully designed considering maximum active power injection capability of each DFIG and the converter overload limits. As demonstrated by the simulation results, the proposed approach not only improves the overall frequency response but also ensures the rotor speed is kept within a safe range. Rotor limits can be maintained without additional actuator for rotor speed protection.  


\bibliographystyle{unsrt}
\bibliography{bibfile}

\end{document}